\documentclass[12pt]{article}
\usepackage{epsfig}
\def\beq{\begin{equation}}
\def\eeq#1{\label{#1}\end{equation}}
\def\eeqn{\end{equation}}

\def\beqa{\begin{eqnarray}}
\def\eeqa#1{\label{#1}\end{eqnarray}}
\def\eeqan{\end{eqnarray}}

\let\bar=\overbar

\def\Dslash{\not{\hbox{\kern-4pt $D$}}}
\def\dslash{\not{\hbox{\kern-2pt $\del$}}}

\def\msb{{\bar{\ssstyle M \kern -1pt S}}}

\def\Title#1{\begin{center} {\Large {\bf #1} } \end{center}}

\newcommand{\bit}{\begin{itemize}}
\newcommand{\eit}{\end{itemize}}
\newcommand{\bea}{\begin{eqnarray}}
\newcommand{\eea}{\end{eqnarray}}
\newcommand{\lsim}{\stackrel{<}{\scriptstyle \sim}}

\begin{document}

\Title{CP violation and cosmology
}

\bigskip\bigskip

%+\addtocontents{toc}{{\it D. Reggiano}}
%+\label{ReggianoStart}

\begin{raggedright}  

{\it Alexander Kusenko \index{Kusenko, A.}\\
Department of Physics and Astronomy \\ 
University of California, Los Angeles,  CA 90095-1547\\
and \\
RIKEN BNL Research Center\\ Brookhaven National
Laboratory, Upton, NY 11973
}
\bigskip\bigskip
\end{raggedright}

\section{Introduction}

CP violation may have played a crucial role in the formation of matter in
the universe.  It must have, if the inflationary cosmology is right, thus
ruling out the (very unpalatable) possibility that the baryon asymmetry was
an initial condition of the Big Bang.  If inflation took place, as most of
us believe based on the growing observational evidence, it made the
universe devoid of matter and set the stage for reheating.  The baryon
asymmetry then must have been produced at some later point through
CP-violating processes.

\section{CP violation and electroweak baryogenesis 
} 

That CP, as well as C and B must be broken for baryogenesis to work, was
first noted by Sakharov~\cite{Sakharov}.  In 1967 the only reason for
considering the baryon number violation was theorists' ambitions.  Neither
experiment, nor favored theoretical models supported this hypothesis.  In
addition to breaking the symmetries, one needs an out-of-equilibrium state
of the universe to generate the asymmetry.

The advent of the Standard Model and 't~Hooft's discovery of a baryon
number violating instanton provided a requisite source of $B$ violation.
The Standard Model also incorporates C, P, and CP breaking.  Although the
baryon number violation is highly suppressed in the Standard Model at zero
temperature, Kuzmin, Rubakov, and Shaposhnikov~\cite{krs} realized that at
high temperatures the baryon number violating processes could go
unsuppressed.  In addition, the universe might be out of thermal
equilibrium at the time of the electroweak phase transition.  Hence,
Kuzmin, Rubakov, and Shaposhnikov~\cite{krs} put forth a very 
plausible and appealing possibility that the baryon asymmetry might arise
from the Standard Model physics, at the time of the electroweak phase
transition~\cite{rs_review}.

Unfortunately, in the minimal Standard Model, this scenario does not work.
First, the phase transition is too weak, unless the Higgs mass is $M_{_H}<
45$~GeV, which is ruled out by experiment.  Second, the CP violation from
the CKM matrix makes a vanishing contribution to baryon asymmetry because
it is suppressed by a product of Yukawa couplings~\cite{bsw}.

Baryogenesis in the supersymmetric extensions of the Standard Model is less
problematic because the additional scalar states can provide both the new
sources of CP violation~\cite{dine,ckn_spont,ckn_susy,cline-kimmo} and a
way to make the phase transition more strongly
first-order~\cite{loops,carena,cline-moore}.  The phase transition is
stronger if one of the stops is very light~\cite{carena}.  The new sources
of CP violation may come, for example, from the chargino mass matrix: 

\begin{equation}
  \overline\psi_R M_\chi \psi_L = (\overline{\widetilde w^{^+}},\
  \overline{\widetilde h^{^+}_{2}} )_{R}
  \left(\begin{array}{cc}
             m_2 & g H_2(x) \\
           g H_1(x) & \mu
        \end{array}\right)
  \left(\begin{array}{c}
         \widetilde w^{^+} \\
         \widetilde h^{^+}_{1}
        \end{array}
  \right)_{\!\!L} + {\rm h.c.} 
\label{mp}  
\end{equation}

\begin{figure}
\centerline{\epsfysize=3.5in \epsfbox{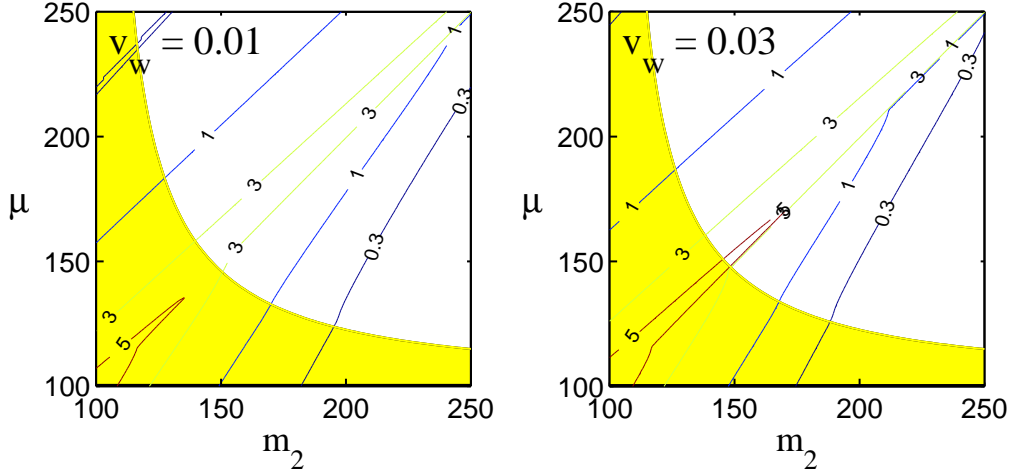}}
\caption{Contours of constant baryon asymmetry in units $10^{-10}$ with
$\sin \delta_\mu = 1$ for (a) $v_w = 0.01$ and (b) $v_w = 0.03$. Mass units
are GeV$/c^2$. Shaded regions are excluded by the LEP2 limit on the
chargino mass, $m_{\chi^\pm}>104$ GeV$/c^2$. To maximize the baryon
asymmetry, one assumes that $\tan \beta \lsim 3$ and that the bubble wall
is very narrow, $\ell_w \simeq 6/T$. From J.M.~Cline et al.~\cite{cjk}.}
\label{fig1} 
\end{figure}

As long as $m_2$ and $\mu$ are complex, spatially varying phases in the
bubble wall provide a source of (spontaneous) CP
violation~\cite{lw,ckn_spont}.  The remaining window for electroweak
baryogenesis in the MSSM is very narrow~\cite{cjk}; several parameters must
be adjusted to maximize the resulting baryon asymmetry (in particular, one
must assume that the wall is very thin, take $\tan \beta <3$, and 
choose the ``optimal'' bubble wall velocity $v_w \approx 0.02$), as shown
in Fig.~\ref{fig1}. 

\begin{figure}
\centerline{\epsfysize=2.5in \epsfbox{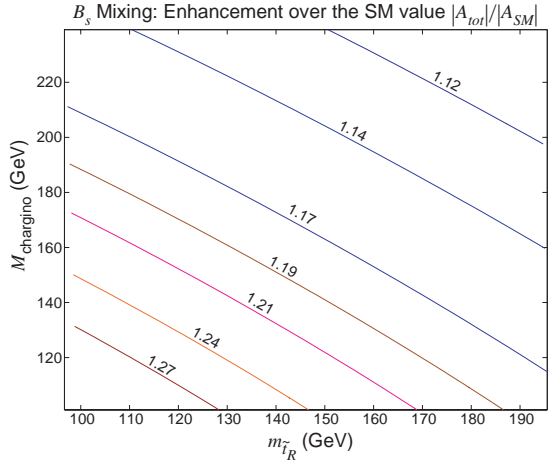}}
\caption{Enhancement of $B_{s}$ in the MSSM, for parameters
  consistent with electroweak baryogenesis~\cite{mp}.  
}
\label{fig2} 
\end{figure}

CP violation from the chargino sector~(\ref{mp}) may enhance the $B_s$
mixing as compared to the Standard Model value~\cite{mp}, especially if the
stop is light (Fig.~\ref{fig2}).  In practice, however, this effect is
observable only if the CKM matrix elements are known to a very high
precision.  In particular, one would need to reduce theoretical
uncertainties in $V_{ub}$ to 5-10\% and in $\sin 2 \beta$ to a few percent.

\section{CP violation and leptogenesis}

If a lepton asymmetry of the universe formed after inflation but before the
electroweak phase transition, the sphalerons, which violate $B$ and $L$ but
preserve  $(B-L)$, would convert (roughly, a half of)
the lepton asymmetry into a baryon asymmetry.  This observation gave rise
to an extremely appealing scenario of leptogenesis~\cite{leptogenesis}.
The relevant CP violation may reside in the neutrino mass matrix, which, in
general, has a number of complex phases~\cite{nu_cp, nu_cp_review}.

\section{Transient CP violation, baryogenesis, and dark matter}

In addition to CP violation hard-wired into the lagrangian, some
manifestations of CP non-conservation may occur for a short period of time
in the early universe, during which time the baryon asymmetry might have
formed.  A well-know example is the Affleck-Dine scenario for
baryogenesis~\cite{ad,drt}, in which CP violating seeds may be effectively
amplified by the motion of the scalar condensate.  Similarly, electroweak
baryogenesis at preheating~\cite{kt,ggks,ck,cgk} can take advantage of
CP-violating motions of time-dependent condensates during
preheating~\cite{cgk}.  CP violation of this kind is poorly constrained by
experiment because it becomes small after thermalization.

\subsection{Affleck-Dine baryogenesis and dark-matter-genesis} 

In models with low-energy supersymmetry inflation can lead to formation of
the Affleck-Dine condensate~\cite{ad,drt} with a large VEV.  A high-scale
physics undoubtedly violates B and CP through higher-dimension operators.
Hence, the motion of the scalar condensate after inflation is not B and CP
symmetric.  Thus, the universe acquires a baryon asymmetry.  The
Affleck-Dine scenario~\cite{ad} is simple, appealing, and flexible in that
the final baryon asymmetry can easily be made consistent with the data.

In addition, the Affleck-Dine baryogenesis can produce dark matter as
well.  In general, the Affleck-Dine condensate does not remain
homogeneous and can break up~\cite{dm} into SUSY Q-balls~\cite{ak_mssm}.
This affords a number of interesting possibilities for generating baryons
and dark matter simultaneously~\cite{dm,em,kk}.

Since baryons and dark matter arise form the same physical process, 

\centerline{\epsfysize=1.5in \epsfbox{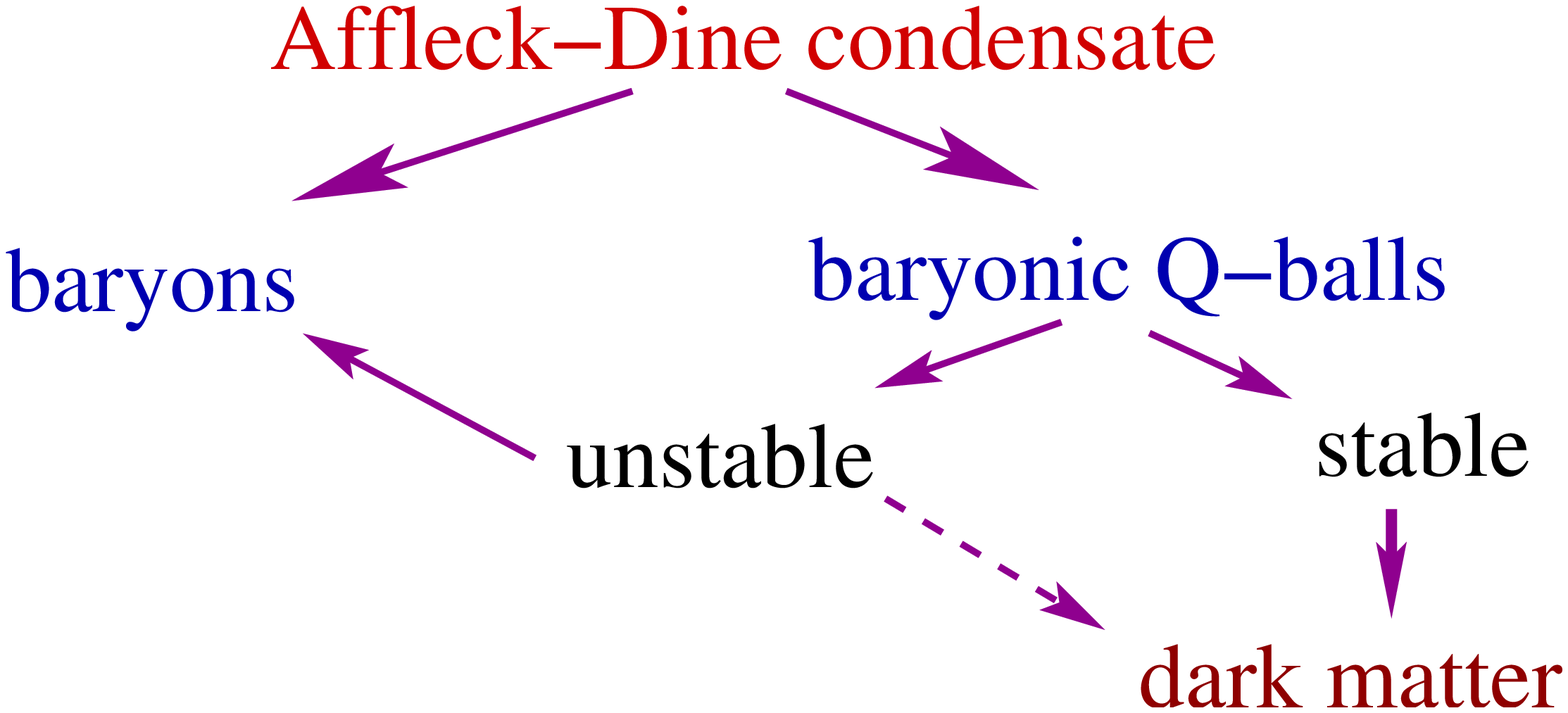}}

\noindent the
ratio of $\Omega_{\rm dark}$ to $\Omega_B$ may have a natural explanation
in some models~\cite{dm_ratio}.

\subsection{Electroweak baryogenesis at preheating}
Several viable scenarios for electroweak baryogenesis at preheating were
presented in Ref.~\cite{cgk}.  For example, a modified spontaneous
baryogenesis a la Cohen, Kaplan, and Nelson~\cite{ckn_spont} becomes very
efficient at preheating.  Their original scenario used the variation of the
Higgs field inside a wall of a bubble formed in a first-order phase
transition.  A similar effect can occur at preheating uniformly in space,
on the horizon scales~\cite{cgk}.  One can obtain the desired baryon
asymmetry in a Standard Model supplemented by an additional Higgs doublet
and an inflaton sector~\cite{cgk}.  The difference with the scenario
proposed by Cohen, Kaplan and Nelson~\cite{ckn_spont} is that in our case
CP violation occurs homogeneously in space, not only inside a bubble
wall. In addition, the final prediction for the baryon asymmetry in the CKN
scenario was very far from the equilibrium value because the sphaleron rate
was slow on the time scales associated with the growth of bubbles.  In the
case of preheating, the Higgs parameters change slowly in time while the
baryon number non-conservation is rapid.  This allows a slow adiabatic
adjustment of the baryon number to that which minimizes the free energy.

Several additional sources of CP violation might affect the physics of
preheating and  facilitate baryogenesis~\cite{cgk}.

\section{Strong CP violation and the axion cosmology} 

The QCD vacuum is a superposition $ |\theta \rangle=\sum_n \exp \{-i n
\theta \} | n \rangle$  of topologically distinct vacuum states $| n
\rangle$. As a result, the QCD Lagrangian can be written as 

\begin{equation}
{\cal L}_{QCD} = {\cal L}_{\rm pert} + \bar{\theta} \frac{g^2}{32 \pi^2} F
\tilde{F},   
\end{equation}
where

\begin{equation}
\bar{\theta} = \theta + \arg \det M
\end{equation}

\begin{figure}
\centerline{\epsfysize=3in \epsfbox{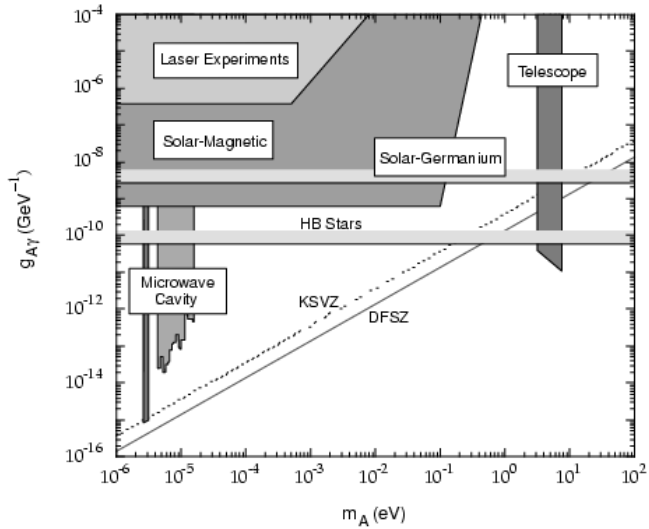}}
\caption{
Exclusion regions for the axion mass vs. coupling, from Particle Data
Book~\cite{pdg}. 
}
\label{fig3}
\end{figure}

Experimentally, the value of $\bar{\theta}$ must vanish to a high
precision, $\bar{\theta} \ll 10^{-10}$.  However, if all the quarks have
non-zero masses, there is no (simple) reason why the phase in the Yukawa
matrix should cancel the QCD vacuum phase, unless $\bar{\theta}$ relaxes to
zero dynamically, by a VEV of a scalar field.  This elegant solution to the
strong CP problem was proposed by Peccei and Quinn~\cite{pq}.  The breaking
of a global U(1) symmetry gives rise to a light scalar field, the
axion~\cite{axion}.  

Several models can accommodate an axion consistent with the existing
experimental bounds \cite{axion_review}.  A light, weakly interacting axion
makes a good candidate for dark matter.  The present experimental limits 
are shown in (Fig.~\ref{fig3}).

\section{Conclusions}

There is every reason to believe that matter-antimatter asymmetry is a
consequence of CP non-conservation in particle physics.  On the other hand,
CP violation from the quark mixing is not sufficient for baryogenesis.
This implies the existence of new, yet undiscovered, sources of CP
violation in nature.

\section{Acknowledgments} 

The author thanks J.~Cline and H.~Murayama for helpful discussions.  This
work was supported in part by the DOE grant DE-FG03-91ER40662.

%\def\Discussion{
%\setlength{\parskip}{0.3cm}\setlength{\parindent}{0.0cm}
%     \bigskip\bigskip      {\Large {\bf Discussion}} \bigskip}
%\def\speaker#1{{\bf #1:}\ }
%\def\endDiscussion{}

%\Discussion
%
%\speaker{D. Giovanni (University of Seville)}  My analysis indicates that the
%recovery of the two gentlemen is due simply to their embrace of the masculine
%principle and has nothing to do with magnetism at all.  Could you comment on 
%this?

%\speaker{Reggiano} Professor Giovanni has discussed this hypothesis in several
%forums, but, I do not believe there is anything in print.  I understand that
%he is spending his time in other pursuits.
%
%\speaker{D. Anna (University of Seville)}  In fact, my colleague Giovanni 
%has expressed opposite opinions on this question at various times, depending
%on the audience.  All of these testosterone-based theories are, of course,
%nonsense.
%
%\endDiscussion
 

\begin{thebibliography}{99}

%%
%%  bibliographic items can be constructed using the LaTeX format in SPIRES:
%%    see    http://www.slac.stanford.edu/spires/hep/latex.html
%%  SPIRES will also supply the CITATION line information; please include it.
%%

%\bibitem{Mesmer}
%F. A. Mesmer, Proc. Wien. Acad. Sci. {\bf 13}, 1564, 1593 (1762).
%%CITATION = PWASA,13,1564;%%


%\cite{Sakharov:dj}
\bibitem{Sakharov}
A.~D.~Sakharov,
%``Violation Of CP Invariance, C Asymmetry, And Baryon Asymmetry Of The
%Universe,'' 
Sm Zh.\ Eksp.\ Teor.\ Fiz.\  {\bf 5} (1967) 32 
%[JETP Lett.\  {\bf 5} (1967\ SOPUA,34,392-393.1991\ UFNAA,161,61-64.1991)
%24] 
.
%%CITATION = ZFPRA,5,32;%%

\bibitem{krs} V.~A.~Kuzmin, V.~A.~Rubakov and M.~E.~Shaposhnikov,
%``On The Anomalous Electroweak Baryon Number Nonconservation In The Early
%Universe,'' 
Phys.\ Lett.\ B {\bf 155}, 36 (1985).
%%CITATION = PHLTA,B155,36;%%


\bibitem{rs_review}
For review, see, {\em e.g.}, V.~A.~Rubakov and M.~E.~Shaposhnikov,
%``Electroweak baryon number non-conservation in the early universe and in
%high-energy collisions,'' 
Usp.\ Fiz.\ Nauk {\bf 166}, 493 (1996)
[Phys.\ Usp.\  {\bf 39}, 461 (1996)]
%[arXiv:hep-ph/9603208]
;
%%CITATION = HEP-PH 9603208;%%
A.~G.~Cohen, D.~B.~Kaplan and A.~E.~Nelson,
%``Progress in electroweak baryogenesis,''
Ann.\ Rev.\ Nucl.\ Part.\ Sci.\  {\bf 43}, 27 (1993)
%[arXiv:hep-ph/9302210].
%%CITATION = HEP-PH 9302210;%%


\bibitem{cjk}
J.~M.~Cline, M.~Joyce and K.~Kainulainen,
%``Supersymmetric electroweak baryogenesis in the WKB approximation,''
Phys.\ Lett.\ B {\bf 417}, 79 (1998); {\it ibid.},\ B {\bf 448}, 321 (1999);  
%[arXiv:hep-ph/9708393]; 
%%CITATION = HEP-PH 9708393;%%
%\cite{Cline:2000nw}
%``Supersymmetric electroweak baryogenesis,''
JHEP {\bf 0007}, 018 (2000).
%[arXiv:hep-ph/0006119]; 
%%CITATION = HEP-PH 0006119;%%
%\cite{Cline:2001rk}
%``Erratum for 'Supersymmetric electroweak baryogenesis',''
%arXiv:hep-ph/0110031.
%%CITATION = HEP-PH 0110031;%%

\bibitem{bsw}
S.~M.~Barr, G.~Segre and H.~A.~Weldon,
%``The Magnitude Of The Cosmological Baryon Asymmetry,''
Phys.\ Rev.\ D {\bf 20}, 2494 (1979).
%%CITATION = PHRVA,D20,2494;%%

\bibitem{dine}
M.~Dine, P.~Huet, R.~J.~Singleton and L.~Susskind,
%``Creating The Baryon Asymmetry At The Electroweak Phase Transition,''
Phys.\ Lett.\ B {\bf 257}, 351 (1991); 
%%CITATION = PHLTA,B257,351;%%
M.~Dine, P.~Huet and R.~J.~Singleton,
%``Baryogenesis at the electroweak scale,''
Nucl.\ Phys.\ B {\bf 375}, 625 (1992).
%%CITATION = NUPHA,B375,625;%%

\bibitem{ckn_spont}
A.~G.~Cohen, D.~B.~Kaplan and A.~E.~Nelson,
%``Spontaneous baryogenesis at the weak phase transition,''
Phys.\ Lett.\ B {\bf 263}, 86 (1991).
%%CITATION = PHLTA,B263,86;%%

\bibitem{ckn_susy}
A.~G.~Cohen and A.~E.~Nelson,
%``Supersymmetric baryogenesis,''
Phys.\ Lett.\ B {\bf 297}, 111 (1992).
%[arXiv:hep-ph/9209245].
%%CITATION = HEP-PH 9209245;%%

\bibitem{cline-kimmo}
J.~M.~Cline, K.~Kainulainen and A.~P.~Vischer,
%``Dynamics of two Higgs doublet CP violation and baryogenesis at the
%electroweak phase transition,'' 
Phys.\ Rev.\ D {\bf 54}, 2451 (1996). 
%[arXiv:hep-ph/9506284].
%%CITATION = HEP-PH 9506284;%%

\bibitem{lw}  T.~D.~Lee, Phys. Rev. D {\bf 8}, 1226 (1973); Phys. Reports
{\bf 9}, 143 (1974); S.~Weinberg, Phys. Rev. Lett. {\bf 37}, 657 (1976).
%%CITATION = PHRVA,D8,1226;%%
%%CITATION = PRPLC,9,143;%%
%%CITATION = PRLTA,37,657;%%



\bibitem{loops} 
J.~R.~Espinosa,
%``Dominant Two-Loop Corrections to the MSSM Finite Temperature Effective
%Potential,'' 
Nucl.\ Phys.\ B {\bf 475}, 273 (1996) %[arXiv:hep-ph/9604320]
; 
%%CITATION = HEP-PH 9604320;%%
D.~Bodeker, P.~John, M.~Laine and M.~G.~Schmidt,
%``The 2-loop MSSM finite temperature effective potential with stop
%condensation,'' 
Nucl.\ Phys.\ B {\bf 497}, 387 (1997). 
%[arXiv:hep-ph/9612364].
%%CITATION = HEP-PH 9612364;%%


\bibitem{carena}
M.~Carena, M.~Quiros and C.~E.~Wagner,
%``Opening the Window for Electroweak Baryogenesis,''
Phys.\ Lett.\ B {\bf 380}, 81 (1996); Nucl.\ Phys.\ B {\bf 524}, 3 (1998). 
%[arXiv:hep-ph/9710401].
%%CITATION = HEP-PH 9710401;%%
%[arXiv:hep-ph/9603420].
%%CITATION = HEP-PH 9603420;%%

\bibitem{cline-moore}
J.~M.~Cline and G.~D.~Moore,
%``Supersymmetric electroweak phase transition: Baryogenesis versus
%experimental constraints,'' 
Phys.\ Rev.\ Lett.\  {\bf 81}, 3315 (1998). 
%[arXiv:hep-ph/9806354].
%%CITATION = HEP-PH 9806354;%%

\bibitem{mp}
H.~Murayama and A.~Pierce,
%``Signatures of baryogenesis in the MSSM,''
hep-ph/0201261.
%%CITATION = HEP-PH 0201261;%%

\bibitem{leptogenesis}
M.~Fukugita and T.~Yanagida,
%``Baryogenesis Without Grand Unification,''
Phys.\ Lett.\ B {\bf 174}, 45 (1986).
%%CITATION = PHLTA,B174,45;%%

\bibitem{nu_cp}
L.~Wolfenstein,
%``Oscillations Among Three Neutrino Types And CP Violation,''
Phys.\ Rev.\ D {\bf 18} (1978) 958; 
%%CITATION = PHRVA,D18,958;%%
V.~D.~Barger, K.~Whisnant and R.~J.~Phillips,
%``CP Violation In Three Neutrino Oscillations,''
Phys.\ Rev.\ Lett.\  {\bf 45}, 2084 (1980); 
%%CITATION = PRLTA,45,2084;%%
G.~C.~Branco, L.~Lavoura and M.~N.~Rebelo,
%``Majorana Neutrinos And CP Violation In The Leptonic Sector,''
Phys.\ Lett.\ B {\bf 180}, 264 (1986); 
%%CITATION = PHLTA,B180,264;%%
A.~Kusenko and R.~Shrock,
%``General determination of phases in leptonic mass matrices,''
Phys.\ Lett.\ B {\bf 323}, 18 (1994); hep-ph/9403315; 
%%CITATION = HEP-PH 9311307;%%
%%CITATION = HEP-PH 9403315;%%
G.~C.~Branco, T.~Morozumi, B.~M.~Nobre and M.~N.~Rebelo,
%``A bridge between CP violation at low energies and leptogenesis,''
Nucl.\ Phys.\ B {\bf 617}, 475 (2001);
%%CITATION = HEP-PH 0107164;%%
W.~Buchmuller and D.~Wyler,
%``CP violation, neutrino mixing and the baryon asymmetry,''
Phys.\ Lett.\ B {\bf 521}, 291 (2001); 
%%CITATION = HEP-PH 0108216;%%
H.~Fritzsch and Z.~z.~Xing,
%``How to describe neutrino mixing and CP violation,''
Phys.\ Lett.\ B {\bf 517}, 363 (2001);
%%CITATION = HEP-PH 0103242;%
G.~C.~Branco, R.~Gonzalez Felipe, F.~R.~Joaquim and M.~N.~Rebelo,
%``Leptogenesis, CP violation and neutrino data: What can we learn?,''
hep-ph/0202030; 
%%CITATION = HEP-PH 0202030;%%
M.~Frigerio and A.~Y.~Smirnov,
%``Structure of neutrino mass matrix and CP violation,''
hep-ph/0202247; 
%%CITATION = HEP-PH 0202247;%%
D.~Chang, A.~Masiero and H.~Murayama,
%``Neutrino mixing and large CP violation in B physics,''
hep-ph/0205111; 
%%CITATION = HEP-PH 0205111;%%
G.~Altarelli and F.~Feruglio,
%``Theoretical models of neutrino masses and mixings,''
hep-ph/0206077. 
%%CITATION = HEP-PH 0206077;%%

\bibitem{nu_cp_review} B. Kayser, these Proceedings. 


\bibitem{ad} 
I.~Affleck and M.~Dine,
%``A New Mechanism For Baryogenesis,''
Nucl.\ Phys.\ B {\bf 249}, 361 (1985).
%%CITATION = NUPHA,B249,361;%%

\bibitem{drt} M.~Dine, L.~Randall and S.~Thomas, Nucl. Phys. {\bf B458}
  (1996) 291; R.~Allahverdi, B.~A.~Campbell and J.~R.~Ellis,
%``Reheating and supersymmetric flat-direction baryogenesis,''
Nucl.\ Phys.\ B {\bf 579}, 355 (2000)
%[arXiv:hep-ph/0001122].
%%CITATION = HEP-PH 0001122;%% A.~Anisimov and M.~Dine,
%``Some issues in flat direction baryogenesis,''
Nucl.\ Phys.\ B {\bf 619}, 729 (2001). 
%[arXiv:hep-ph/0008058].
%%CITATION = HEP-PH 0008058;%%

\bibitem{kt}  L.~M.~Krauss and M.~Trodden, Phys. Rev. Lett.{\bf 83}, 1502
(1999). 
%%CITATION = HEP-PH 9902420;%%

\bibitem{ggks}  J.~Garc\'ia-Bellido, D.~Grigoriev, A.~Kusenko, and
M.~Shaposhnikov, Phys. Rev. D {\bf 60}, 123504 (1999).  
%%CITATION = HEP-PH 9902449;%%


\bibitem{ck}  J.~M.~Cornwall and A.~Kusenko,
%``Baryon number non-conservation and phase transitions at preheating,''
Phys.\ Rev.\ D {\bf 61}, 103510 (2000)
%%CITATION = HEP-PH 0001058;%%
J.~Garcia-Bellido and D.~Y.~Grigoriev,
%``Inflaton-induced sphaleron transitions,''
JHEP {\bf 0001}, 017 (2000). 
%[arXiv:hep-ph/9912515].
%%CITATION = HEP-PH 9912515;%%

\bibitem{cgk} J.~M.~Cornwall, D.~Grigoriev and A.~Kusenko,
%``Resonant amplification of electroweak baryogenesis at preheating,''
Phys.\ Rev.\ D {\bf 64}, 123518 (2001)
%%CITATION = HEP-PH 0106127;%%


\bibitem{dm} 
A.~Kusenko and M.~E.~Shaposhnikov,
%``Supersymmetric Q-balls as dark matter,''
Phys.\ Lett.\ B {\bf 418}, 46 (1998); 
%[arXiv:hep-ph/9709492].
%%CITATION = HEP-PH 9709492;%%
K.~Enqvist and J.~McDonald,
%``Q-balls and baryogenesis in the MSSM,''
Phys.\ Lett.\ B {\bf 425}, 309 (1998)
%[arXiv:hep-ph/9711514].
%%CITATION = HEP-PH 9711514;%%

\bibitem{ak_mssm} A.~Kusenko,
%``Solitons in the supersymmetric extensions of the standard model,''
Phys.\ Lett.\ B {\bf 405}, 108 (1997); 
%%CITATION = HEP-PH 9704273;%%
Phys.\ Lett.\ B {\bf 404}, 285 (1997);  
%%CITATION = HEP-TH 9704073;%%
A.~Kusenko, V.~Kuzmin, M.~E.~Shaposhnikov and P.~G.~Tinyakov,
%``Experimental signatures of supersymmetric dark-matter Q-balls,''
Phys.\ Rev.\ Lett.\  {\bf 80}, 3185 (1998).
%%CITATION = HEP-PH 9712212;%%

\bibitem{em} K.~Enqvist, J.~McDonald: 
%``Q-balls and baryogenesis in the MSSM,''
Phys.\ Lett.\  B {\bf 425}, 309 (1998); 
%%CITATION = HEP-PH 9711514;%%
%%CITATION = HEP-PH 9803380;%%
%``D-term inflation and B-ball baryogenesis,''
Phys.\ Rev.\ Lett.\  {\bf 81}, 3071 (1998);
%%CITATION = HEP-PH 9806213;%%
%``MSSM dark matter constraints and decaying B-balls,''
Phys.\ Lett.\  B {\bf 440}, 59 (1998);
%%CITATION = HEP-PH 9807269;%%
%K.~Enqvist and J.~McDonald,
%``Observable isocurvature fluctuations from the Affleck-Dine condensate,''
Phys.\ Rev.\ Lett.\  {\bf 83}, 2510 (1999); 
%%CITATION = HEP-PH 9811412;%%
%``The dynamics of Affleck-Dine condensate collapse,''
%%CITATION = HEP-PH 9908316;%%
M.~Fujii and K.~Hamaguchi,
%``Higgsino and wino dark matter from Q-ball decay,''
Phys.\ Lett.\ B {\bf 525}, 143 (2002);
%[arXiv:hep-ph/0110072].
%%CITATION = HEP-PH 0110072;%%
M.~Fujii and K.~Hamaguchi,
%``Non-thermal dark matter via Affleck-Dine baryogenesis and its detection
%possibility,'' 
hep-ph/0205044.
%%CITATION = HEP-PH 0205044;%%


\bibitem{kk} 
S.~Kasuya and M.~Kawasaki,
%``Q-ball formation: Obstacle to Affleck-Dine baryogenesis in the
%gauge-mediated SUSY breaking?,''
Phys.\ Rev.\ D {\bf 61}, 041301 (2000);
%%CITATION = HEP-PH 9909509;%%
Phys.\ Rev.\ D {\bf 62}, 023512 (2000); 
%%CITATION = HEP-PH 0002285;%%
Phys.\ Rev.\ D {\bf 64}, 123515 (2001). 
%[arXiv:hep-ph/0106119].
%%CITATION = HEP-PH 0106119;%%
S.~Kasuya,
%``Difficulty of a spinning complex scalar field to be dark energy,''
Phys.\ Lett.\ B {\bf 515}, 121 (2001). 
%[arXiv:astro-ph/0105408].
%%CITATION = ASTRO-PH 0105408;%%

\bibitem{dm_ratio}
K.~Enqvist and J.~McDonald,
%``B-ball baryogenesis and the baryon to dark matter ratio,''
Nucl.\ Phys.\ B {\bf 538}, 321 (1999); 
%[arXiv:hep-ph/9803380].
%%CITATION = HEP-PH 9803380;%%
M.~Laine and M.~E.~Shaposhnikov,
%``Thermodynamics of non-topological solitons,''
Nucl.\ Phys.\ B {\bf 532}, 376 (1998); 
%[arXiv:hep-ph/9804237].
%%CITATION = HEP-PH 9804237;%%
M.~Fujii and T.~Yanagida,
%``A solution to the coincidence puzzle of Omega(B) and Omega(DM),''
hep-ph/0206066.
%%CITATION = HEP-PH 0206066;%%


\bibitem{pq}
R.~D.~Peccei and H.~R.~Quinn,
%``CP Conservation In The Presence Of Instantons,''
Phys.\ Rev.\ Lett.\  {\bf 38}, 1440 (1977); 
%%CITATION = PRLTA,38,1440;%%
%``Constraints Imposed By CP Conservation In The Presence Of Instantons,''
Phys.\ Rev.\ D {\bf 16}, 1791 (1977).
%%CITATION = PHRVA,D16,1791;%%

\bibitem{axion} 
S.~Weinberg,
%``A New Light Boson?,''
Phys.\ Rev.\ Lett.\  {\bf 40}, 223 (1978); 
%%CITATION = PRLTA,40,223;%%
F.~Wilczek,
%``Problem Of Strong P And T Invariance In The Presence Of Instantons,''
Phys.\ Rev.\ Lett.\  {\bf 40}, 279 (1978).
%%CITATION = PRLTA,40,279;%%

\bibitem{axion_review} For review, see, {\em e.g.}, 
R.~D.~Peccei,
%``Particle Physics Footprints Of The Invisible Axion,''
Phys.\ Scripta {\bf T36}, 218 (1991); 
%%CITATION = PHSTB,T36,218;%%
J.~E.~Kim,
%``A theoretical review of axion,''
astro-ph/0002193; 
%%CITATION = ASTRO-PH 0002193;%%
K.~van Bibber and D.~Kinion,
%``Review Of Dark-Matter Axion Experiments,''
Nucl.\ Phys.\ Proc.\ Suppl.\  {\bf 91} (2001) 376.
%%CITATION = NUPHZ,91,376;%%

\bibitem{pdg} D.E. Groom et al., Europ. Phys. J., C15 (2000) 1. 

\end{thebibliography}
\end{document}